\documentclass[psfig]{aa}
\begin{document}

\title{The formation of helium lines in the spectrum of COM J1740-5340}


\author{L.B.Lucy}

\offprints{L.B.Lucy}

\institute{Astrophysics Group, Blackett Laboratory, Imperial College 
London, Prince Consort Road, London SW7 2AZ}

\date{Received ; accepted }

\maketitle

\begin{abstract}

The He$\:${\sc i} $\lambda$5876\AA\ absorption line recently discovered in 
the spectrum of the companion to the millisecond pulsar PSR J1740-5340  
is tentatively attributed to electron impact excitations due to the
irradiation of its atmosphere
by $\gamma$-rays emitted by the pulsar's magnetosphere. Numerical
calculations, similar to those carried out previously for Type Ib SNe,
indicate that a pulsar beam with photon energies $\sim$ 1 MeV gives rise
to a $\lambda$5876\AA\ line of the observed strength if the beam's
spin-down conversion efficiency approaches 1\%.
However, a significant difficulty for the proposed mechanism is the strength of
the singlet line at $\lambda$6678\AA. Compared to the corresponding triplets,
singlet lines are weak because of the loss of excitation when
photons emitted in the series $ n^{1}P^{0} \rightarrow 1^{1}S$   
ionize hydrogen atoms, an effect absent in the hydrogen-free atmospheres of
Type Ib SNe.


\end{abstract}

\section{Introduction}

The binary millisecond pulsar PSR J1740-5340 
was recently discovered  in the globular cluster NGC 6397 by D'Amico et al.
(2001a) with the Parkes radio telescope. They reported a 
spin period of $3.65$ms, a $1.35$ day circular orbit, and
eclipses at $1.4$ GHz that last for more than 40\% of the orbital period.
Subsequent Parkes observations (D'Amico et al. 2001b) allowed an accurate     
timing solution yielding the spin-down rate as well as the 
binary's coordinates. 

	With the position known, Ferraro et al. (2001) used
archived images to obtain a secure optical identification   
with a star (WF4-1) previously listed as a BY Dra candidate
(Taylor et al. 2001). 
The pulsar's companion (named COM J1740-5340 by Ferraro et al. 2003)
has a luminosity close to that
of the cluster's main-sequence turnoff stars but its colour is anomalously
red. In
addition, the companion is variable with a range of $\sim 0.2$mag. This
variability has the same period as the binary orbit
and matches that due to the tidal distortion of a star filling
its Roche lobe. Ferraro et al. (2001) emphasize that, in this regard,
PSR J1740-5340 differs from two other eclipsing millisecond pulsars, where 
strong photometric variability is clearly due to the heating of 
the companions' facing hemispheres by the pulsars' emission.
The absence of this expected heating effect is also
reported by Kaluzny et al. (2003) and by Orosz \& van Kerkwijk (2003). 

	Although not apparent in broad-band light curves, the effect of
the pulsar's emission on its companion has almost certainly been
detected spectroscopically by Ferraro et al. (2003). 
From observations made 
to determine the spectroscopic orbit, these authors made
the serendipitous discovery of He$\:${\sc i} absorption lines at
5876 and 6678\AA. Given that the
companion's effective temperature T$_{eff} \simeq 5530$K 
(Ferraro et al. 2003) is substantially less
than the T$_{eff}$ 
($ \ga 10,000$K) at which the He$\:${\sc i} lines are present for main
sequence stars, this discovery is a major surprise and obviously significant. 

	Ferraro et al. (2003) explain the presence of He$\:${\sc i}
absorption in terms of a narrow strip across the illuminated face of the
companion within which the local
T$_{eff} \sim 10,000$K due to heating by beamed pulsar radiation. But 
outside this strip, irradiation is negligible and so the local
T$_{eff}$ is then $\sim 5000K$. By a suitable choice of parameters, 
Ferraro et al. imply that the observed strengths of the $\lambda$5876\AA\
line can reconciled with the absence of a photometric heating 
effect. In a very recent paper, this group (Sabbi et al. 2003) have published
more of their spectroscopy and repeated their conclusion that the pulsar's
emission pattern must be highly anisotropic. 

	Although the model proposed
by Ferraro et al. is not implausible, another effect suggests itself if we
note that COM J1740-5340 has two points in common with Type Ib SNe.
The first is that the spectra of these SNe are also remarkable in
exhibiting  He$\:${\sc i} 
absorption lines despite cool photospheric continua 
(Harkness et al. 1987). 

	The second point in common - the presence of high-energy photons - 
follows from the explanation for the helium lines in the Ib's. In the
standard interpretation (Lucy 1991, hereafter Paper I; Swartz et al. 1993 ),
a huge
overpopulation of the $n=2$
levels is caused by impact excitations and ionizations of helium atoms
by non-thermal
electrons. These energetic electrons are the cascade products created by the
multiple Compton scatterings of $\gamma$-rays emitted by the radioactive
nuclei $^{56}$Co. There is a significant energy density of such
$\gamma$-rays in the reversing layers of these SNe because of mass loss
preceding the explosion and the mixing out of 
$^{56}$Ni following the explosion.

	In view of these similarities, an obvious conjecture is that 
$\gamma$-rays emitted by the magnetosphere of PSR J1740-5340, 
or arising when the pulsar's relativistic wind encounters the companion's
wind, irradiate the atmosphere of its companion, overpopulate the $n=2$
levels of He$^{0}$, and hence give rise to detectable
absorption lines. The investigation of this conjecture is the subject of this
paper.

\section{An irradiated model atmosphere}   

	In this section, the problem is defined and then the steps needed to
calculate the He line spectrum are briefly described. The discussion
follows closely that given in Paper I for Type Ib SNe.

\subsection{The problem}   

	In investigating the non-thermal excitation of
helium in the atmosphere of COM J1740-5340, we face
several major uncertainties. First and foremost, there is no direct
evidence that PSR J1740-5340 emits a powerful beam of hard radiation.
Although the object is detected in X-rays (Grindlay et al. 2001, 2002),
its X-ray luminosity $L_{X}$ is
weaker by a factor $\sim 17$ than expected for its spin-down
luminosity $\dot{E}$ 
from the correlation
$L_{X} \sim 10^{-3}\dot{E}$ found for millisecond pulsars in the field
(Becker \& Tr\"umper 1997). However, the detected X-rays are probably
not emitted from
the pulsar's magnetosphere but rather from the interaction region between
the pulsar's relativistic wind and the mass loss from the companion
(Grindlay et al. 2002).

	Despite this absence of evidence for a beam of hard radiation, we
assume that such a beam exists
and that it sweeps over at least part of the companion's disk.
Accordingly, the problem investigated in this paper is to study the 
formation of the He$\:${\sc i} spectrum at a typical {\em point} on the
companion's disk lying within the strip swept out by the beam.

	The problem as posed refers to irradiation by a pencil beam emitted
by the pulsar's magnetosphere. But the calculation is also relevant to
irradiation by an extended source of hard radiation, such as that due to
the colliding winds of the pulsar and its companion. The
single beam then represents a one-point quadrature approximation for the
effect of the extended source.

\subsection{The incident beam}   

	Because everything about the beam is conjectural, we take it to be
mono-energetic, comprising photons of single energy $\epsilon_{0}$, 
the beam's effective photon energy.
If the beam's flux as a function $\epsilon$ were known, $\epsilon_{0}$
could be defined as the value of $\epsilon$ such that the mono-energetic
beam deposits the same fraction of its flux in the reversing layer
as would the actual beam. Nevertheless, since
the vertical distribution of the deposited energy would not be exact,
the concept of an effective photon energy is not rigorous.

	The phase-averaged flux incident on the companion's atmosphere is
conveniently written as
\begin{equation}
 {\cal F}_{0} = \mu_{0} \times \eta \times \frac{\dot{E}}{4 \pi a^{2}}
           \times \frac{4 \pi}{\Omega}
\end{equation}
where  $\mu_{0}$ is the incident direction cosine,
$a$ is the orbital
separation, $\eta$ is the conversion efficiency of $\dot{E}$ into beamed
radiation, and $\Omega$ is the solid angle swept out by the beam.

	The adopted numerical values are as follows:
$\dot{E} = 1.4 \times 10^{35}$ erg s$^{-1}$ (D'Amico et al. 2001b);
$a = 4.33  \times 10^{11}$ cm , corresponding to component masses
${\cal M}_{1}$ = 1.5 ${\cal M}_{\sun}$, 
${\cal M}_{2}$ = 0.26 ${\cal M}_{\sun}$ (Ferraro et al. 2003); and
$\Omega = 1$ steradian. This last choice is standard practice in view of
our ignorance concerning beaming fractions.

	Inserting these values, we have 
${\cal F}_{0} = 7.48 \times 10^{11} \mu_{0}\eta$ erg cm$^{-2}$ s$^{-1}$.

\subsection{Model atmosphere}   

A full NLTE investigation of the response of the companion's atmosphere 
to irradiation by a beam of $\gamma$ rays is not
justified in view of the above-mentioned uncertainties. Accordingly,
the approach adopted is to embed a NLTE calculation for He atoms in an
atmosphere that is otherwise unperturbed by irradiation. Moreover, a
severely simplified model is adopted for this unperturbed atmosphere.
Specifically, a plane-parallel, Schuster-Schwarzschild model is assumed
comprising a black-body emitting photosphere at Rosseland mean
optical depth $\tau = 2/3$ and an isothermal reversing layer
with temperature $T = 0.93T_{eff}$. On the assumption of hydrostatic 
equilibrium and constant Rosseland mean opacity $\kappa_{R}$,  
the variation of gas pressure in the reversing layer
is given by $P = \tau \times g/\kappa_{R} $, where $g$ is the surface gravity.
The adopted constant opacity $\kappa_{R}$ is the solution of the equation
$ P = 2/3 \times g/\kappa_{R} (P,T_{eff})$. 

	The parameters of the model atmosphere are
$T_{eff} = 5530K$  (Ferraro et al. 2003), $log g = 3.44$, X=0.75 and Y=0.25. 
The value of $log g$ corresponds to radius $R_{2} = 1.6 R_{\sun}$
(Orosz \& van Kerkwijk 2003) and mass ${\cal M}_{2}$ = 0.26 ${\cal M}_{\sun}$.
The adopted chemical composition neglects metals
since [Fe/H]$\sim -2$ (Ferraro et al.) and assumes that envelope stripping
has not yet exposed He-enriched zones.

	With these parameters, the equation for the opacity has the
solution $\kappa_{R} = -1.13$ dex. This was obtained by iterative
interpolation
in Table 6 of Alexander \& Ferguson (1994). Although this Table is computed
for $Z = 0.02$, the H$^{-}$ ion
is the dominant source of opacity for the given photospheric parameters,
with metals as minor contributors to the electron density $n_{e}$.

	In the subsequent calculations, the reversing layer is discretized
into 50 layers of width $\Delta log \tau = 0.1$, within each of which physical
variables are constant. The resulting small optical depths of the outermost
layers ensure that line centres are accurately calculated.   

\subsection{Beam loss rates}   

The first step in calculating the He spectrum is to derive the rate at which
the beam deposits energy in the reversing layer in the form of
energetic electrons. Each incident $\gamma$-ray creates a cascade of
non-thermal electrons by multiple Compton scatterings off free and bound
electrons.

	At optical depth $\tau$, the rate at which
$\gamma$-rays are losing energy by Compton scattering is 
\begin{equation}
 {\cal L}(\tau) = 4 \pi \int f_{\epsilon} \sigma_{\epsilon}
           J_{\epsilon} d\epsilon  
\end{equation}
where $\sigma_{\epsilon}$ is the Compton scattering coefficient,
$f_{\epsilon}$ is
the expected fraction of $\epsilon$ transferred to the Compton electron, and
$J_{\epsilon}(\tau)$ is the mean intensity at photon energy $\epsilon$. 
Note that here, and throughout the paper, rates refer to unit volume.

	As in Paper I, a Monte Carlo code is used to simulate this
process. If the incident flux ${\cal F}_{0}$ at $\tau = 0$ is represented by
${\cal N}$
identical photon packets containing $\gamma$-rays of energy $\epsilon_{0}$ and
entering the atmosphere in time interval $\Delta t$, then each
packet's initial energy $w_{0}$ is given by
${\cal N} \times w_{0}/\Delta t = {\cal F}_{0}$. These packets are
followed as they propagate within the reversing layer until they escape
at $\tau = 0$ or penetrate below the photosphere at $\tau = 2/3$, where they
are assumed to thermalize.
 
	When the scattering histories of all ${\cal N}$ packets are
complete, the loss rate in the $m$-th layer is derived with
the Monte Carlo estimator (Lucy 1999, 2003)
\begin{equation}
 {\cal L}_{m} = \frac{w_{0}}{\Delta t} \frac{1}{V}
   \sum \frac{w_{\epsilon}}{w_{0}} \: \Delta s \:f_{\epsilon} \:
   \sigma_{\epsilon}
\end{equation}
Here $V$ is the volume of a vertical column of unit cross section
through the $m$-th layer, and $\Delta s$ is the pathlength of a packet of
energy $w_{\epsilon}$ between consecutive events, where an event is either a
Compton scattering or a crossing of a layer's boundary. The summation is over 
all pathlengths in layer $m$.

	Monte Carlo convergence experiments show
that this estimator gives fractional errors $\la 10^{-3}$ for
${\cal N} = 10^{6}$, the sample size adopted here. Note that this estimator
returns a non-zero value for the loss rate even if no $\gamma$-ray packet
undergoes a Compton scattering in layer $m$. 

	Because ${\cal L}$ is proportional to $\eta$, it is
convenient to
compute its value ${\cal L}^{1}$ for $\eta = 1$. The energy
deposition rate in
layer $m$ is then
$\eta \: {\cal L}^{1}_{m}(\epsilon_{0},\mu_{0})$.

\subsection{Deposition fractions}

	The next step is to determine the rates of impact excitations 
and ionizations of He atoms.
To achieve this, we must follow the slowing down of fast
electrons as they interact with matter in the reversing layer and record
the energy transfers into the various deposition channels.

	This calculation is greatly facilitated by two approximitions  
adopted in Paper I. First, noting that the stopping distance for an
MeV electron is $\sim 10^{-2}$ that for an MeV photon (Colgate, Petschek \&
Kriese 1980), we assume that the electrons deposit all their
energy in situ. The second approximation (Meyerott 1978) makes use of the
asymptotic
near constancy of $D_{k}(E)$, the fraction of an electron's kinetic 
energy $E$ that is deposited in channel $k$.
Thus we need only compute the
deposition fractions $D_{k}^{\dag}$ for $E^{\dag} = 10$keV since 
$D_{k}(E) \simeq D_{k}^{\dag}$ for $E > E^{\dag}$. 

	The values $D_{k}^{\dag}$ are obtained by solving the Spencer-Fano
equation - Eq (2) in Paper I. This requires that we identify the stopping
mechanisms and provide their cross sections. In Paper I, this equation was
solved for a weakly-ionized pure helium gas, so that the stopping 
mechanisms were impact excitations and ionizations of He$^{0}$ atoms
and Coulomb scattering with free (thermal) electrons. Now the equation must
be solved for a weakly-ionized H-He gas, so the stopping mechanisms to be
added are impact excitations and ionizations of H atoms.

	The treatment of He impacts and Coulomb scattering remain as
described in
Sect. 2.3 of Paper I. The added impact cross sections for 
H are the approximate formulae of Johnson (1972), with the
H atom represented by 14 bound levels with principal quantum numbers
$n = 1-14$. In addition, the probability distribution for the energy $E_{s}$
of the secondary electron when H is ionized is taken to be 
$ \propto (E_{s}^{2}+J^{2})^{-1}$ with $J = 8$eV (Shull 1979).

	With composition specified, the deposition fractions $D_{k}^{\dag}$
are functions of just one variable $x_{e}$, the fractional ionization of H.
In conformity with the assumption (Sect. 2.3) that the atmosphere
is unperturbed by irradiation, electron densities are computed from Saha's
equation. Then, neglecting the variation with height,
we set $x_{e} = x_{e}^{*} = 5.2 \times 10^{-5}$, its value at $\tau = 1/3$.
With this approximation, the deposition of non-thermal energy throughout the
reversing layer is 
governed by the single vector $D_{k}^{\dag}(x_{e}^{*})$.

	Summing the 
appropriate elements of this vector, we find that 11.7\%  of the deposited
energy is converted directly into heat via Coulomb scattering, 82.0\% 
is converted into ionization and excitation energy of H, and the remaining
6.3\% is converted into ionization and excitation energy of He.

\subsection{Statistical equilibrium of helium}   

	The rate at which non-thermal energy is absorbed in exciting He atoms
to level $i$ in layer $m$ is
$\eta \: {\cal L}^{1}_{m}  \times D_{i}^{\dag} $.
Accordingly, if $W_{i}$ is the energy of this level
relative to the ground state, the rate of impact excitation (ionization) 
of level $i$ in layer $m$ is
\begin{equation}
   \Gamma_{i} = \eta \times
   {\cal L}^{1}_{m} (\epsilon_{0},\mu_{0})  \times \frac{D_{i}^{\dag}}{W_{i}}
\end{equation}

	If $\Lambda_{ji} n_{j}$ denotes the rate at which transitions
$j \rightarrow i$ occur due to conventional radiative and collisional
(thermal) processes,
then, with impacts included, the equations of statistical equilibrium for
He levels $i = 2,3, \ldots, \kappa$ are    
\begin{equation}
 \sum_{j = 1}^{\kappa} (\Lambda_{ji}n_{j}- \Lambda_{ij}n_{i}) = -  \Gamma_{i}
\end{equation}
When the constraint $\sum_{1}^{\kappa} n_{i} = n_{He}$ is added, we have
$\kappa$ equations in the $\kappa$ unknowns $n_{i}$. This system must be
solved
in each layer of the atmosphere in order to compute the emergent profiles of
the He lines. But since the radiative coefficients in Eq. (5) depend
on the solutions of the line transfer problems, iterations are required
(Sect. 2.8).

	The data sources for the $\Lambda$ coefficients in Eq. (5) are as
stated in Paper I, except that the collision strengths of transitions between
the lowest five levels of He have been updated using the calculations of
Sawey \& Berrington (1993). 

\subsection{Escape probabilities}   

	For most of the He lines, we can assume that 
line photons in the reversing layer interact only with their own transition -
i.e., continuum
process can be neglected. But this is not true for transitions to the ground
state since the emitted photons can ionize hydrogen and the atmosphere is
optically thick in the Lyman continuum. 

	For the adopted atomic model, this is relevant for the
permitted ground state transitions $ n^{1}P^{0} \rightarrow 1^{1}S$ at 584,
537 and 522\AA. Since the 
optical depths for these transitions are also huge, we can assume 
that when such a line photon is emitted it 
is absorbed in situ either by the line itself or by a H atom.
Accordingly, we adopt an escape probability formalism for these three
transitions and only consider thermal motions in computing Doppler broadening.

	If $\phi_{\nu}$ denotes the normalized absorption profile, a line
photon emitted with frequency $\nu$ sees a line
absorption coefficient $\ell_{\nu} = a_{\nu} n_{j}$, where   
\begin{equation}
  a_{\nu} = \frac{\pi e^{2}}{m_{e} c} \: f_{ji} \: 
   \left(1- \frac{g_{j} n_{i}}{g_{i} n_{j}} \right) \: \phi_{\nu}
\end{equation}
and a continuum absorption coefficient  $ k_{\nu} = a_{H}(\nu) \: n_{H}$,
where we assume all H atoms are neutral and in the ground state and neglect
the correction for stimulated emission. The probability that this particular
photon
will be lost to the line is therefore $ k_{\nu}/( k_{\nu} +  \ell_{\nu})$.
Thus, if we average over the emission profile, the escape probability
is
\begin{equation}
  \beta_{ij} = \int \frac{k_{\nu}}{k_{\nu} + \ell_{\nu}} \: \phi_{\nu} \:
 d \nu 
\end{equation}
Note that the emission profile is taken to be identical to the absorption
profile on the assumption of complete redistribution (Mihalas 1978, p.29).

	This treatment is applied to the permitted transitions 
$i \rightarrow j$ for $j= 1$. As a result, the radiative contribution of these
transitions to Eq. (5) simplifies to 
$A_{ij} \: \beta_{ij} n_{i}$, where $A_{ij}$ is the Einstein coefficient for 
spontaneous emission.

	For the isothermally stratified reversing layer, these escape
probabilities are independent of $\tau$. With $\phi_{\nu}$ given by the Voigt
function
assuming pure radiation damping, we obtain $\beta_{ij}$ =
$1.8 \times 10^{-3},\; 4.4 \times 10^{-3}$ and $9.2  \times 10^{-3}$ for
the lines at 584, 537 and 522\AA, respectively.   

	An escape probability treatment is also adopted for recombinations to
the ground state. The emitted photon  will travel a negligible distance before
it photoionizes a H or a He atom. Since the recombination photons are
strongly concentrated to the He ionization threshold at $\nu_{T}$, the
escape probability is accurately given by (Osterbrock 1974, p.23)
\begin{equation}
  \beta_{\kappa 1} = \frac{n_{H} a_{H}(\nu_{T})}
                         {n_{H} a_{H}(\nu_{T}) +  n_{He} a_{He}(\nu_{T})}
\end{equation}
Accordingly, if $\alpha_{1}$ denotes the total recombination coefficient for
the ground state, it is replaced in the equations of statistical equilibrium
by $\beta_{\kappa 1} \alpha_{1}$. With the He threshold
cross section $7.48 \times 10^{-18}$cm$^{2}$ of Fernley et al. (1987), we
obtain $\beta_{\kappa 1} = 0.68$

\subsection{Line formation}   

	Subordinate transitions are not treated with escape probabilities
but by solving the transfer equation. For such transitions, the radiative
contribution to Eq. (5)
is $ A_{ij} n_{i} - (B_{ji} n_{j} - B_{ij} n_{i}) \bar{J}_{ij}$, where
\begin{equation}
     \bar{J}_{ij}    = \int J_{\nu} \: \phi_{\nu} \: d \nu 
\end{equation}
is the profile-averaged mean intensity. This quantity is
obtained by numerical integration after solving
the line transfer equation
\begin{equation}
  \mu \: \frac{\partial I_{\nu}}{\partial z } = - \ell_{\nu} \: (I_{\nu} - S) 
\end{equation}
with source function
\begin{equation}
  S = \frac{2h \nu^{3}}{c^{2}} /
    \left( \frac{g_{i}n_{j}}{g_{j}n_{i} } - 1 \right) 
\end{equation}
for a grid of frequencies spanning the transition. Since high precision is
not required, Eq. (10) is solved only at $\mu = \pm 1/2$ giving
$J_{\nu} = (I_{\nu}^{+}+I_{\nu}^{-})/2$. The boundary conditions on the
inward and outward beams are $I_{\nu}^{-} = 0$ at $\tau = 0$ and
$I_{\nu}^{+} = B_{\nu} (T_{eff})$ at  $\tau = 2/3$.

	As in Sect. 2.7, the line profile $\phi_{\nu}$ is given by the Voigt
function assuming pure radiation damping. But in view of the moderate
optical depths of the subordinate transitions, a microturbulent contribution
of 3 km s$^{-1}$ is now included in the Doppler parameter.

	A simple iterative procedure is adopted to obtain mutually consistent
solutions of the equations of statistical equilibrium and line transfer.
Eq (5) is first solved with  $\bar{J}_{ij} =   B_{\nu} (T_{eff})/2$, its
optically-thin limit. The resulting level populations are then
used to obtain improved values of $\bar{J}_{ij}$ from the transfer equation.
This process is repeated until the average of the absolute fractional changes
of the level populations in the atmosphere are $ < 10^{-6}$. 

	When the iterations have converged, each line's emergent
flux is $F_{\nu} (0) = I^{+}_{\nu}(0)$, from which its residual intensity
$r_{\nu} = F_{\nu} (0)/ B_{\nu} (T_{eff})$ and equivalent width 
$W_{\lambda} = \int (1-r_{\nu}) d \lambda$ are readily computed.

\subsection{Hydrogen lines}

	With only minor modifications, the code described above can also
be used to compute the H line spectrum of the irradiated atmosphere of
COM J1740-5340. The vector $D_{k}^{\dag}(x_{e}^{*})$ already contains
elements giving the fractions of $\eta \: {\cal L}^{1}_{m}$
used to excite and ionize H atoms. Accordingly, with the standard assumption
of in situ re-absorption of Lyman- continuum and line photons, the
H line spectrum is similarly obtained by iteratively solving the statistical
equilibrium and transfer equations of Sects. 2.6 and 2.8. One point of
difference is that impact ionizations of H are allowed to increase $n_{e}$
from its LTE value.

\section{Numerical results}   

	In this section, the simple treatment of line formation
in an irradiated atmosphere developed in Sect.2 is applied to compute the
strengths of He lines
in the spectrum of COM J1740-5340. With atmospheric parameters fixed
(Sect. 2.3), line strengths depend on $\mu_{0}, \eta$ and $\epsilon_{0}$.
Selecting
a typical point on the disk, we set $\mu_{0} = 2/3$ and thus limit 
the investigation to exploring the atmosphere's response to a beam with
conversion
efficiency $\eta$ and effective photon energy $\epsilon_{0}$.     

\subsection{Deposition efficiency}   

	Line strengths depend on $\epsilon_{0}$ only through the function 
${\cal L}^{1}_{m}(\epsilon_{0},\mu_{0})$ defined in Sect. 2.4. From the
Monte Carlo calculations made in determining this function, it is informative
to compute the fraction of incident flux 
deposited in the reversing layer, the fraction deposited below the photosphere
at $\tau = 2/3$, and the fraction emerging
back into space at $\tau = 0$.

	Fig. 1 plots these deposited fractions as a function of $\epsilon_{0}$
for $\mu_{0} = 2/3$. The behaviour of these curves is readily understood
from the physics of Compton scattering. At low energies, the cross section
is approximately that of Thomson scattering and
a negligible fraction of the photon's energy is transferred to 
the Compton electron. As a result, most of the incident flux is simply
reflected back into space, with only a small fraction deposited within the
reversing layer.

	At high energies, on the other hand, the greatly reduced Klein-
Nishina cross section
allows a large fraction of the incident photons to pass right through the
reversing layer, so that most of the incident flux is thermalized below the
photosphere. The local $T_{eff}$ is therefore increased and, in consequence,
so also are temperatures in the reversing layer. In this way,
a sufficiently strong beam of $\gamma$-rays with
$\epsilon_{0}(MeV) \ga 30 $ 
has an {\em indirect} effect on the observed line spectrum, with
the $n = 2$ levels of He$^{0}$ being populated by the photospheric continuum
and by collisions with thermal electrons. This is the 
circumstance envisaged by Ferraro et al. (2003).

	However, for a broad range of intermediate photon energies
$ 0.04  \la  \epsilon_{0}(MeV) \la 10  $, we see from the bell-shaped curve
in Fig. 1 that
a significant fraction of the incident flux is deposited in the reversing
layer ($\tau < 2/3$) and thus {\em directly} affects line
formation through impact excitations and ionizations.

\begin{figure}
\vspace{8.2cm}
\includegraphics{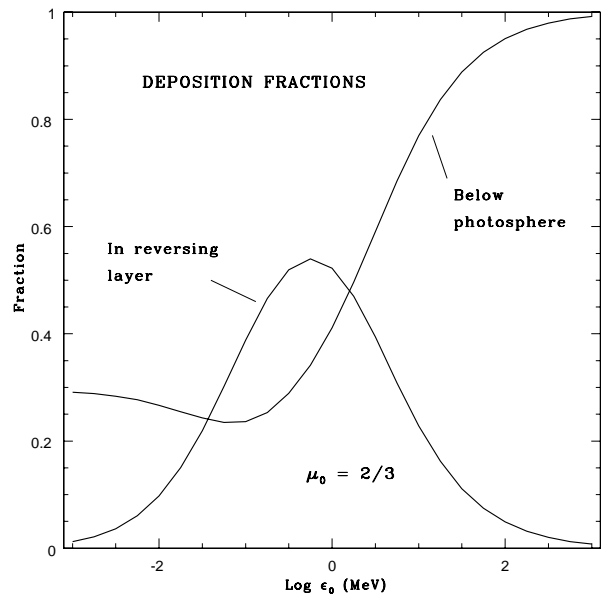}
\caption{Deposition of beam flux. The fractions of incident flux
deposited above and below the photosphere ($\tau = 2/3$) are plotted against
the photon
energy $\epsilon_{0}$ of the mono-energetic beam. The incident direction
cosine $\mu_{0} = 2/3$.}
\end{figure}

\subsection{Curves of growth}   

	Fig. 1 shows that non-thermal
excitation of the He$\:${\sc i} lines is favoured when the beam's photon
energy is $\simeq$ 1 Mev. Accordingly, we now set  $\epsilon_{0} =$ 1 MeV,
since a definitive failure at this $\epsilon_{0}$ 
would be fatal for this interpretation.

	With $\epsilon_{0}$ fixed, the single remaining parameter is
$\eta$, the conversion efficiency. We now take $\eta$ to be a
free parameter, postponing till later a discussion of observational and
theoretical constraints on its magnitude (Sect. 4.2). 

	As $\eta$ increases from zero, the impact rates $\Gamma_{i}$ in
Eq.(5) increase proportionately - see Eq.(4), resulting eventually in
column densities for the $n = 2$ levels sufficient for detectable
absorption of the photospheric continuum. The simplest way of illustrating
this effect is
by constructing curves of growth with $\eta$ as the abscissa and
dimensionless equivalent width $W_{\lambda}/\lambda$ as ordinate.
Accordingly,
for $Log \:\eta$ ranging from -5 to -0.5, the coupled radiative transfer
and statistcal
equilibrium problems of Sect. 2 have been solved and the resulting
line profiles and equivalent widths $W_{\lambda}$ computed for all
transitions represented in the adopted atomic model for He$^{0}$.

	Fig. 2 is the curve of growth for the line
He$\:${\sc i} $\lambda$5876\AA. We see that  $W_{\lambda}$ increases 
linearly with $\eta$ for $\eta \la 4 \times 10^{-4}$ and thereafter more
slowly. This is reminiscent of conventional curves of growth, where the linear
part is followed by a flat part as the lines' Doppler cores become saturated
while their damping wings remain optically thin. But here there is
the additional effect of the partial filling-in of the absorption line by
intrinsic
line emission within the reversing layer. These additional line photons
are created in the radiative cascades following impact ionizations
and excitations.

\begin{figure}
\vspace{8.2cm}
\includegraphics{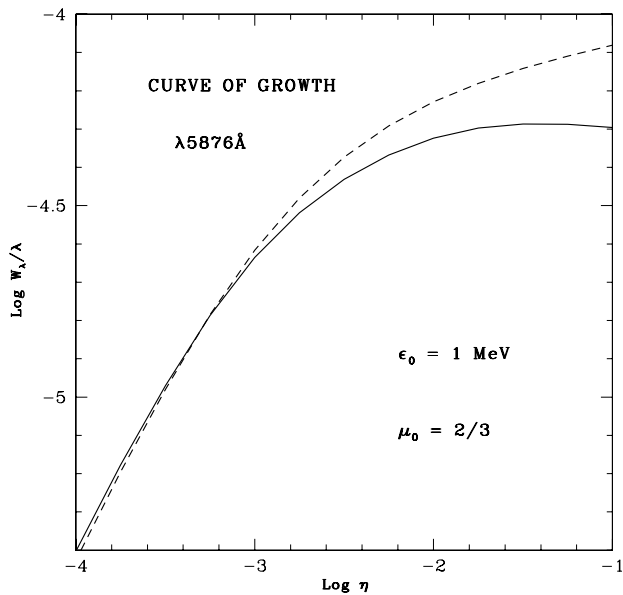}
\caption{Curve of Growth for He$\:${\sc i} $\lambda$5876\AA. The
 dimensionless equivalent width $W_{\lambda}/\lambda$ is plotted against
 the conversion efficiency $\eta$ for the indicated beam parameters. 
 The dashed curve corresponds to line formation by coherent scattering.} 
\end{figure}

	The role of intrinsic emission in reducing $W_{\lambda}$ can be
quantified approximately by also computing, for the given level populations,
the line profile under the assumption of coherent scattering. For the adopted
Schuster-Schwarzschild reversing layer, the residual intensity is then
$r_{\nu} = 1/(1+0.75 N_{j} a_{\nu})$, where $N_{j}$ is the column
density of absorbing atoms and $a_{\nu}$ is given by Eq.(6) . The curve of
growth thus obtained is the dashed
curve in Fig.2. The vertical difference between the two curves increases
with $\eta$, reflecting the growing contribution from cascades.

\begin{figure}
\vspace{8.2cm}
\includegraphics{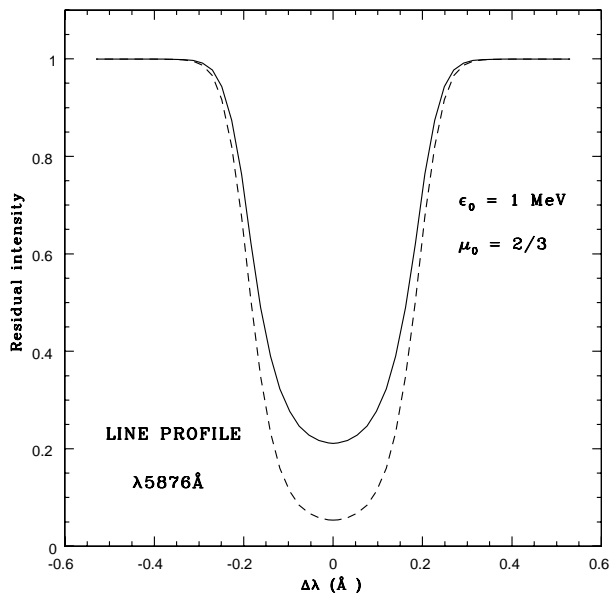}
\caption{Line profile of He$\:${\sc i} $\lambda$5876\AA\ for conversion
 efficiency $\eta = 0.01$. The beam parameters are same as for Fig. 2.
 The dashed curve corresponds to line formation by coherent scattering.} 
\end{figure}

	The actual line profiles used to derive these curves of growth
are plotted in Fig. 3 for $\eta = 10^{-2}$. In this case, intrinsic emission
reduces $W_{\lambda}$ from 0.35 to 0.28\AA.

	Since the strengths of other optical He$\:${\sc i} lines might also be
relevant to testing this excitation mechanism, a selection of their curves of
growth are plotted in Fig.4. As with the $\lambda$5876\AA line,
all lines have curves of growth that eventually depart from linearity in
reponse to the combined effects of core saturation and intrinsic emission. 
The most extreme case is that of the $\lambda$6678\AA\ line, whose 
$W_{\lambda}$ increases to a maximum of $\simeq 0.022$\AA\ at
$\eta \simeq 5 \times 10^{-3}$ and then declines precipitously,
going into emission for $\eta > 1.3 \times 10^{-2}$.

\begin{figure}
\vspace{8.2cm}
\includegraphics{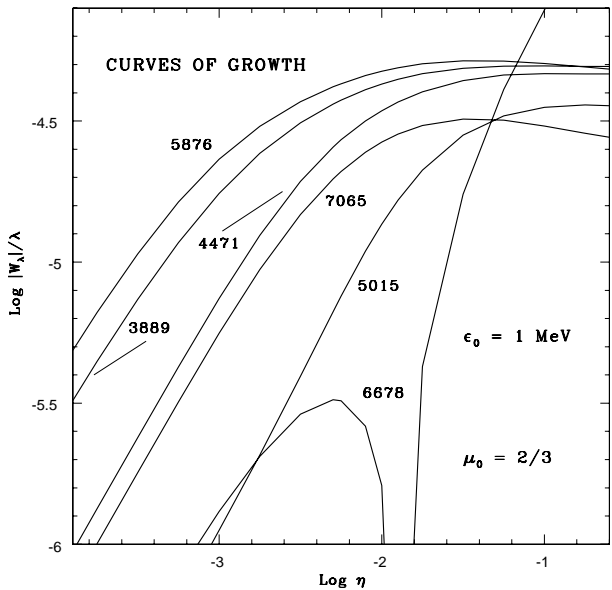}
\caption{Curves of Growth for selected He$\:${\sc i} lines. The
 dimensionless equivalent widths $W_{\lambda}/\lambda$ are plotted against
 the conversion efficiency $\eta$. The beam parameters are same as for
 Fig.2. Note that the line $\lambda$6678\AA\ is in emission for 
 $\eta > 0.013$.} 
\end{figure}

\section{Comparison with observations}   

In this section, the predictions of the non-thermal excitation model
are compared with observational data for  
COM J1740-5340.

\subsection{The equivalent width of $\lambda$5876\AA}   

The prediction that each line has a maximum $W_{\lambda}$ offers an immediate
prospect of observational contradiction. From Fig.2, we see that 
$W_{5876}$ reaches a maximum of $0.30$\AA\ at  $\eta = 4 \times 10^{-2}$. 
For comparison, observed values can be
estimated from the 
spectra published by Sabbi et al. (2003; Fig.4). These spectra show that
$W_{\lambda}$ varies with phase but reaches strengths of $\simeq 0.19$\AA\
at phases 0.02 and 0.56. 
 
	This comparison shows that the predicted maximum exceeds observed
values with a comfortable safety margin. This is just as well since, at these
phases, a significant fraction of the companion's projected disk is not
irradiated by the beam and is thus not expected to contribute to the 
integrated line profile.   

	Eventually, when higher S/N spectra are obtained, it will be
worthwhile to calculate line profiles integrated over the disk. This will not
only give more definitive tests of line strengths but also allow the
phase-dependent velocity displacement of the line centre from the orbital
velocity of
COM J1740-5340 to be tested. Such displacements are a consequence of only one
hemisphere of the synchronously rotating companion (Ferraro et al. 2003) 
being irradiated.
Of course, with data of exceptional quality, the 
technique of Doppler tomography could be used to make inferences about the
pulsar's beam pattern.

\subsection{Conversion efficiency}   

	From Fig.2, we find that $W_{5876}$ is comparable with the observed
strengths for $\eta \ga 2 \times 10^{-3}$. But this assumes the beam sweeps
out solid angle $\Omega = 1$ steradian (Sect. 2.2). This must be increased by
a factor $\ga 3.2$ for the entire facing hemisphere to be irradiated. Thus,
more realistically, the required efficiency is $\eta \ga 6 \times 10^{-3}$. 
The next question, therefore, is whether this is
achievable for photon energies $\sim 1$ MeV.

	No data relevant to this question has yet been acquired for
PSR J1740-5340.
But such data is available for the binary millisecond pulsar
PSR J0218+4232,
whose spin period (2.3ms) and spin-down luminosity ($2.5 \times 10^{35}$
erg s$^{-1}$) (Mineo et al. 2000) are not too different from those of
PSR J1740-5340. Detections and upper limits for photon energies from
0.05 - 10,000 MeV are available for this pulsar. From this data, Kuiper et al.
(2000) suggest that this pulsar's maximum luminosity density is reached in
the COMPTEL MeV range just below the reported upper limits. Accepting this
suggestion and thus interpreting the 2$\sigma$ upper limit reported for
the
COMPTEL 0.75-3 MeV band as a detection, we find $\eta_{obs} \sim$ 15\% for
a 1 steradian beam. This is to be compared with $\eta_{obs} \sim$ 7\%      
found by Kuiper et al. from EGRET {\em detections} for energies between
100 MeV and 1 GeV. 

	This interpretation of the data for PSR J0218+4232 indicates that
the efficiency required if $\epsilon_{0} \sim$1 MeV presents no problem. 
But note that,
for this same data set, Dyks \& Rudak (2002) offer a theoretical
interpretation which predicts fluxes in the COMPTEL bands far below the
reported upper limits. Specifically, at $\sim$1 MeV their model predicts
$\eta \sim 10^{-3}$, corresponding to a barely detectable line at   
$\lambda$5876\AA. 

\subsection{Heating}   

	The 1-steradian efficiency $\eta \sim 2 \times 10^{-3}$ needed to
produce the
observed strength of the $\lambda$5876\AA\ line corresponds to an incident
$\gamma$-ray flux 
${\cal F}_{0} =  7.5 \times 10^{8}$erg cm$^{-2}$ s$^{-1}$.
Even if all of this were thermalized below the photosphere, the increase in
$T_{eff}$ of the irradiated hemisphere is only $\sim 20$K.
Thus the proposed explanation of the $\lambda$5876\AA\ line is consistent
with the absence of a
detectable photometric signature of heating (Sect. 1).  

\subsection{Hydrogen lines}   

	Hydrogen line profiles have also been calculated
as a function of $\eta$. The initial thought was that the
observed strength of the $\lambda$5876\AA\ line might be achieved
simultaneously with H$\alpha$ going into emission, thus explaining 
the star's earlier listing as a BY Dra candidate (Taylor et al. 2001).
In the event, at values of $\eta$ corresponding to $\lambda$5876\AA\
reaching its observed strength, the contribution of intrinsic emission to
H$\alpha$ is still neglible. 

	However, this is not a failure of the model since it is clear
from the spectra published by
Sabbi et al. (2003) that H$\alpha$ emission originating close to the surface
of COM J1740-5340 is {\em not} due to irradiation. This conclusion 
follows because the H$\alpha$ emission remains stong
even when most of the facing hemisphere is hidden from the observer.
Most likely, as Sabbi et al. conclude, the emission has a chromospheric
origin as commonly found in short-period, low mass binaries.
 
\subsection{The equivalent width of $\lambda$6678\AA}   

The above discussion has focussed on He$\:${\sc i} $\lambda$5876\AA\ 
since this the line from which Ferraro et al. (2003) and  
Sabbi et al. (2003) infer that the pulsar's beam is
highly anisotropic.

	Of the other He$\:${\sc i} lines, the one whose
predicted strength is most likely to be in conflict with observation is the
$\lambda$6678\AA\ line, which is the analogue for the singlets of the
triplet line $\lambda$5876\AA. Ferraro et al. (2003) report the presence of
the $\lambda$6678\AA\ line but not its strength.
According to these calculations, its maximum equivalent width is only 
$0.022$\AA, which is probably inconsistent with detetection.

	Subsequent to the original submission of this paper,
E.Sabbi (Bologna) has confirmed that the $\lambda$6678\AA\ line is indeed
reliably detected. Moreover, it reaches an equivalent width well in excess of the above maximum. Specifically, at phase 0.56, she reports that
$W_{6678} \simeq 0.19$ \AA, almost an order of magnitude larger than the predicted maximum.

	The weakness of $\lambda$6678\AA\ in these calculations contrasts 
with its strength in models for Type Ib SNe (Paper I, Fig.1). The origin
of this difference is chemical composition. The atmospheres of Type Ib SNe   
are hydrogen-free, so there is no loss of population from singlet levels due
to the ionization of hydrogen by photons emitted by the
permitted ground state transitions $ n^{1}P^{0} \rightarrow 1^{1}S$ 
(Sect. 2.7). 

	This explanation for the predicted weakness of the
$\lambda$6678\AA\ line is
confirmed by recomputing 
the model at $Log \eta = -2.3$ but now with the line escape probabilities
of Sect. 2.7 $\beta_{i \rightarrow 1} = 0$. The result is a dramatic
increase in $W_{6678}$ from 0.022 to 0.33\AA.

	Interestingly, in this experiment $W_{5876}$ 
also increases, from 0.26 to 0.35\AA. The reason for this is the
removal
of the population decrement of the $n = 2$ singlet levels relative to the
corresponding triplet levels. In consequence, there is no longer 
a significant net collisional transfer of excitation from the triplets
to the singlets.

\subsection{Other lines}   

	 In his report, the referee F.R.Ferraro (Bologna) generously
included unpublished information on other He lines.
This data, as also that from E. Sabbi discussed above, is from the spectra
described by Ferraro et al. (2003) and Sabbi et al. (2003).

	The triplet line $\lambda$3889\AA\ has a predicted strength close
to that of the  $\lambda$5876\AA\ line - see Fig.4 - and so should be seen
were it not blended with the H8 Balmer line. 

	The triplet line $\lambda$4471\AA\ has a predicted equivalent width
of 0.06\AA\ when $W_{5876}$ = 0.19\AA, as observed at phases 0.02
and 0.56. Though weak, a line of this strength should still be visible,
but Ferraro and colleagues finds no evidence of it. This is undoubtedly
a difficulty but one unlikely to be specific to the non-thermal mechanism.
The reason for this is that the $\lambda$5876 and $\lambda$4471\AA\ lines
have the same lower level and the ratio of their predicted strengths   
is approximately equal to the ratio of their f-values. Thus a similar ratio
should be expected for the thermal mechanism proposed by
Ferraro et al. (2003)

\section{Conclusion}   

The aim of this paper has been to investigate the possibility that the
formation mechanism for He$\:${\sc i} absorption lines in the spectrum of 
COM J1740-5340 is similar to that operating in the atmospheres of
Type Ib SNe. This alternative merits serious consideration
in view of the powerful and remarkable constraint on the pulsar's radiation
pattern that follows if the He$\:${\sc i} lines are indeed formed by
conventional thermal processes.

	The result of this investigation is moderately discouraging for
the non-thermal interpretation but not yet a decisive rejection. On the
positive side, the crucial line  
$\lambda$5876\AA\ achieves the observed strength at seemingly feasible
values of the conversion efficiency $\eta$. Moreover, the corresponding
heating effect is consistent with the non-detection of any such effect       
photometrically.

	However, despite these successes, the strengths of other He$\:${\sc i}
lines, especially $\lambda$6678\AA, present difficulties.
The question then is whether such difficulties
are fatal or can be overcome by refinement or modification of the simple
model used here. Among refinements to be considered are the effect of the
beam on the continuum opacity, specifically the dissociation of the H$^{-}$
ion by impacts (Geltman 1960), the ionization of surface layers by
impacts, and the inclusion of an empirical model for the
chromosphere discovered by Sabbi et al. (2003). A higher level treatment
of a stellar atmosphere's response to irradiation by gamma rays is an
attractive technical challenge and likely to have application to other
close companions of exotic objects.  

	Nevertheless, given the relatively low S/N of the spectra of
COM J1740-5340 and
the poorly constrained parameters of the irradiation model, it is perhaps
doubtful that a decision between the thermal and non-thermal line formation
mechanisms can soon be reached purely on the basis of line diagnostics.
Accordingly, the prediction by Sabbi et al. (2003) that up to 1\% of the
disk of COM J1740-5340 radiates with $T_{eff} \sim 10,000$K should be tested
with deep UV imaging at $\lambda \la 2000$\AA. A detection of a photometric
signature of a heating effect at such wavelengths would be a decisive
confirmation of the thermal mechanism and therefore of a narrowly-confined  
pulsar beam.

\acknowledgements

I am indebted to E.Sabbi and the referee, F.R.Ferraro, for unpublished 
data on He line strengths.

\end{document}